\documentclass[11pt]{article} 
\usepackage{fix-cm}
\usepackage{amssymb}
\usepackage{amsmath}
\usepackage{graphicx}
\usepackage{subfigure}
\usepackage{makeidx}
\usepackage{multicol}
\usepackage{natbib}

\newcommand{\Normal}{\mathcal{N}}
\newcommand{\Bino}{\mathcal{Bin}}
\newcommand{\Betadis}{\mathcal{Beta}}

\newcommand{\Prob}{\mathcal{Prob}}
\input CFR.sty  

\begin{document}
\author{Peter J. Green\thanks {School of Mathematics, University of
Bristol, Bristol BS8 1TW, UK.
\newline \hspace*{5mm} Email: {\tt P.J.Green@bristol.ac.uk}.}\\UTS Sydney, Australia \& University of Bristol, UK}
\title{\sc Introduction to finite mixtures\footnote{A chapter prepared for the forthcoming {\it Handbook of Mixture Analysis}}}
\maketitle
\section{Introduction and Motivation}\label{sec:introPG}

Mixture models have been around for over 150 years, as an intuitively simple and practical tool for enriching the collection of probability distributions available for modelling data. In this chapter we describe the basic ideas of the subject, present several alternative representations and perspectives on these models, and discuss some of the elements of inference about the unknowns in the models. Our focus is on the simplest set-up, of finite mixture models, but we discuss also how various simplifying assumptions can be relaxed to generate the rich landscape of modelling and inference ideas traversed in the rest of this book.

\begin{figure}[ht]
\begin{center}
\resizebox{0.9\textwidth}{!}{\includegraphics{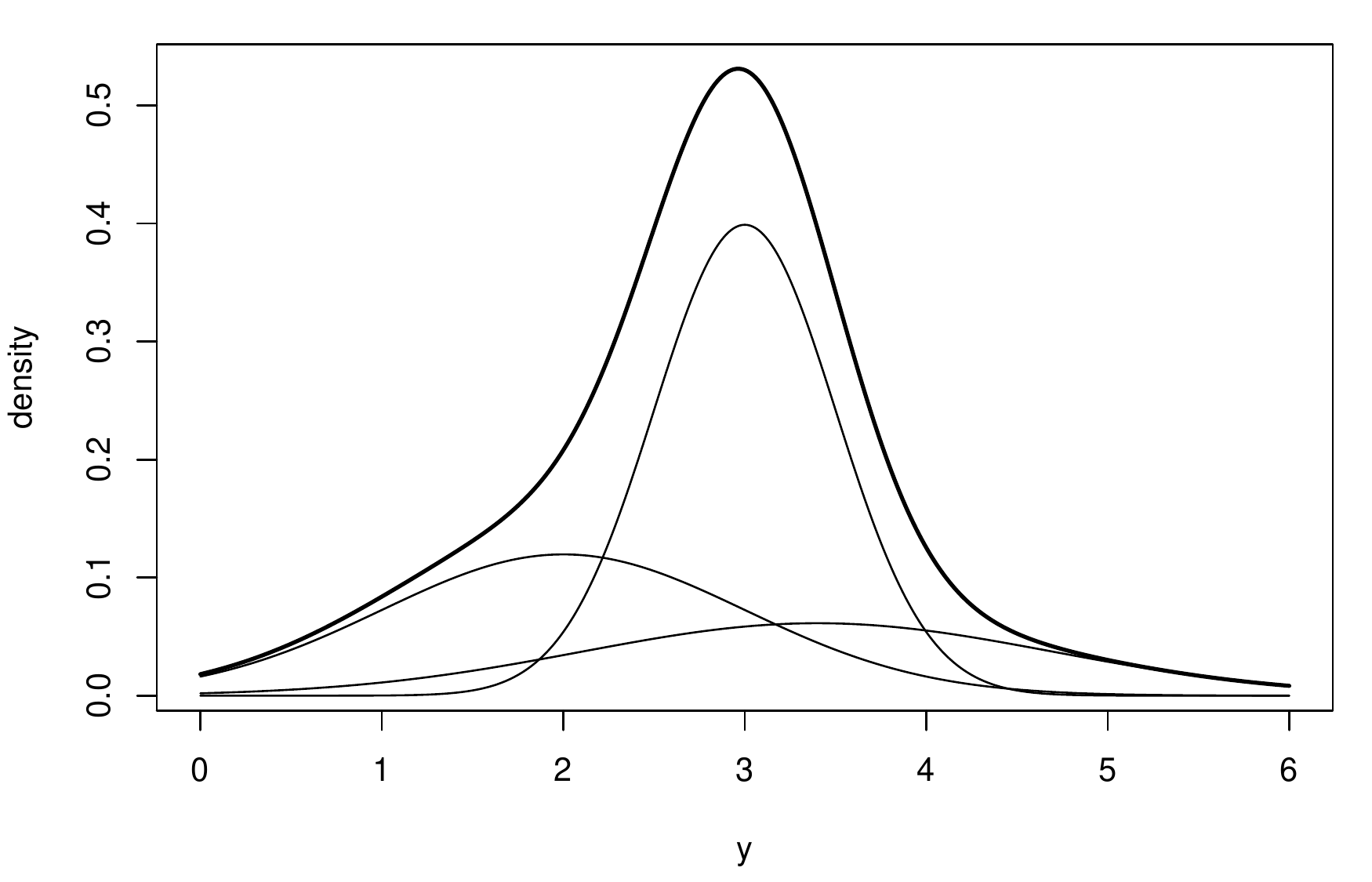}}
\end{center}
\caption{Example of a mixture of 3 Normal densities, giving a unimodal leptokurtic density with slight negative skew. The model is $0.3\times \Normal(2,1)+0.5\times \Normal(3,0.5^2)+0.2\times \Normal(3.4,1.3^2)$.}\label{fig:norm3}
\end{figure}

\begin{figure}[ht]
\begin{center}
\resizebox{0.9\textwidth}{!}{\includegraphics{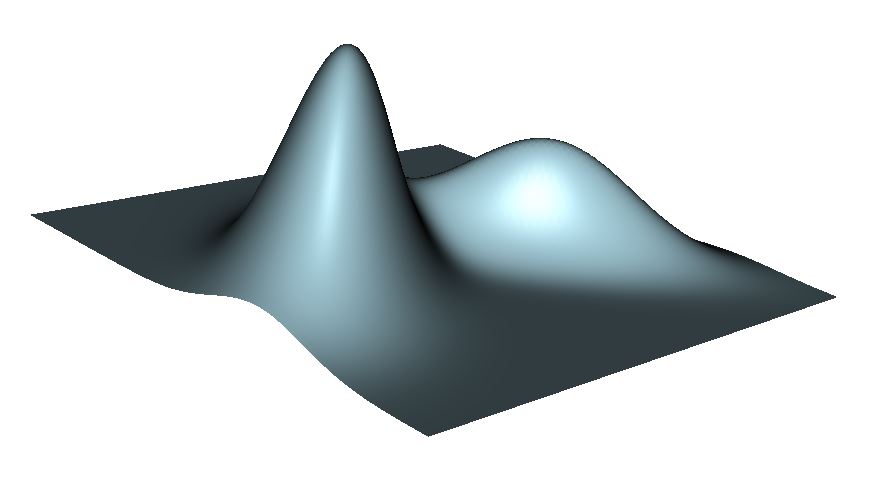}}
\end{center}
\caption[Example of a mixture of 2 bivariate Normal densities.]
{Example of a mixture of 2 bivariate Normal densities. The model is
$0.7\times \Normal\left(\begin{pmatrix} -1 \\ 1\end{pmatrix},\begin{pmatrix} 1 & 0.7 \\ 0.7 & 1\end{pmatrix}\right)
+0.3\times \Normal\left(\begin{pmatrix} 2.5\\0.5\end{pmatrix},\begin{pmatrix} 1 & -0.7 \\ -0.7 & 1\end{pmatrix}\right)$.}
\label{fig:bivnorm}
\end{figure}

\begin{figure}[ht]
\begin{center}
\resizebox{0.9\textwidth}{!}{\includegraphics{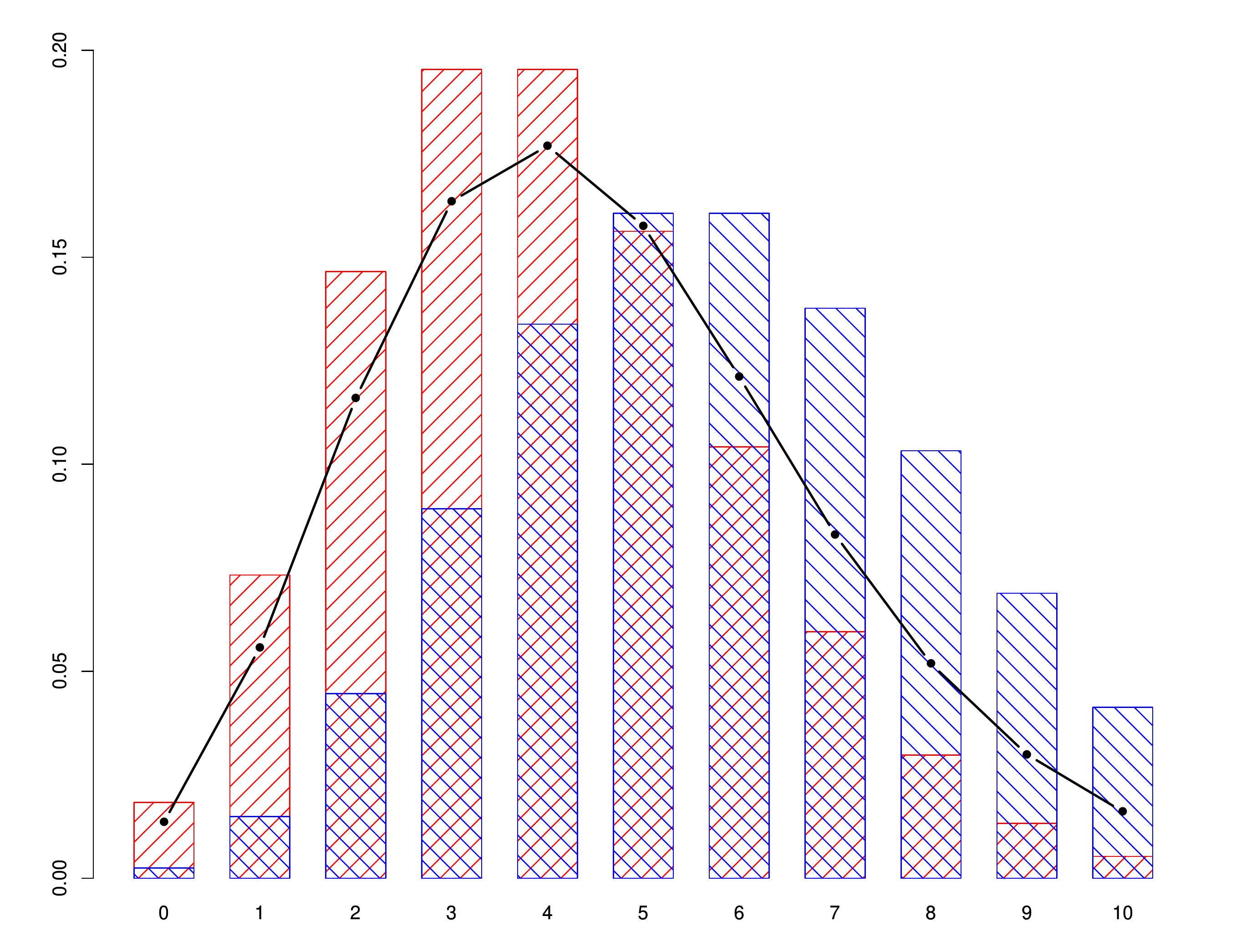}}
\end{center}
\caption{A mixture of two Poisson distributions: $0.7\times\mathcal{P}(4)+0.3\times\mathcal{P}(6)$.}\label{fig:poismix}
\end{figure}

\begin{figure}[ht]
\begin{center}
\resizebox{0.9\textwidth}{!}{\includegraphics{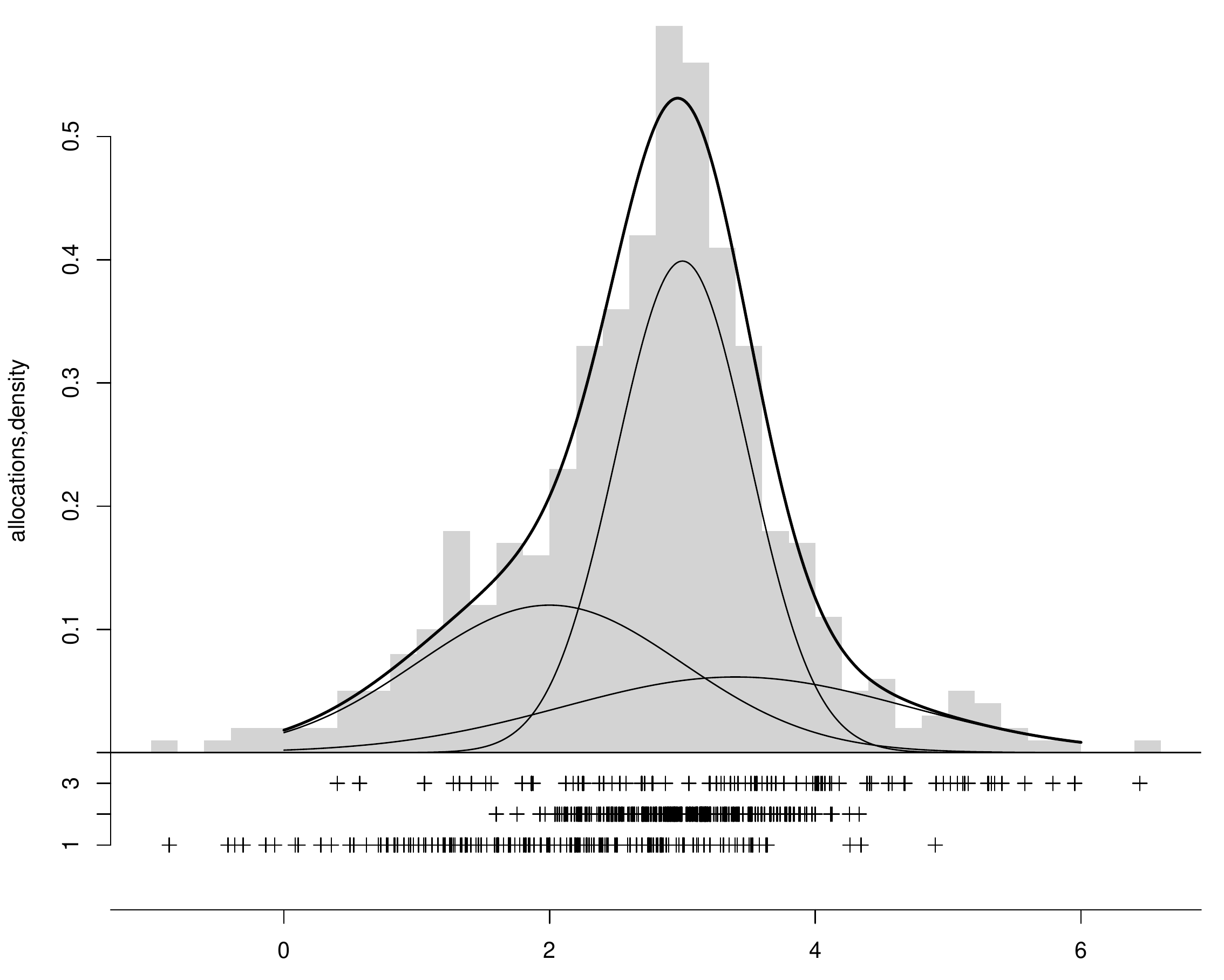}}
\end{center}
\caption{The same mixture of 3 Normal densities as in Fig. \ref{fig:norm3}, together with a sample of size 500, sampled according to (\ref{eq:latent}), histogrammed, and plotted against the allocation variables.}\label{fig:latent}
\end{figure}

\subsection{Basic formulation}

Sometimes a simple picture tells nearly the whole story: the basic finite mixture model is illustrated by Figure \ref{fig:norm3}. It assumes that data $y$ are drawn from a density modelled as a \emph{convex combination} of \emph{components} each of specified parametric form $f$:

\begin{equation}\label{eq:fmm}
y\sim \sum_{g=1}^G \eta_g f(\cdot|\theta_g);
\end{equation}
we use the conditioning bar `$|$' without necessarily implying a Bayesian viewpoint on inference.
The usual set-up is that data are modelled as a simple random sample $\by=(y_1,y_2,\ldots,y_n)$ of variables drawn independently from (\ref{eq:fmm}) but of course this form of density could find a role in many other data-modelling contexts.

In inference about a mixture model, the component density $f$ is fixed and known; the component-specific parameters $\theta_g$ (often vector-valued) and the weights $\eta_g$, non-negative and summing to 1, are usually (but not always) considered to be unknown, and the number of components $G$ is also sometimes unknown. There may be partial equality among $\theta_g$; for example, in the case of a Normal mixture where $f(\cdot)=\phi(\cdot|\mu,\sigma)$, the Normal density, we might sometimes be interested in location mixtures with a constant variance, so $\theta_g=(\mu_g,\sigma)$ or sometimes in a scale mixture $\theta_g=(\mu,\sigma_g)$.

The model (\ref{eq:fmm}) exhibits an inherent \emph{exchangeability} in that it is invariant to permutation of the component labels. This exchangeability is considered by many authors to imply a kind of \emph{unidentifiability}; we discuss this issue below in Section \ref{sec:tech}.

Another, mathematically equivalent, way to write the model (\ref{eq:fmm}) is in integral form,
\begin{equation}\label{eq:cmm}
y\sim \int f(\cdot|\theta)H(d\theta).
\end{equation}
The finite mixture model (\ref{eq:fmm}) corresponds to the discrete case
where the mixing measure $H$ places probability mass $\eta_g$ on the atom $\theta_g$, $g=1,2,\ldots,G$.

It can be useful to allow slightly more generality:
\begin{equation}\label{eq:fmmg}
y\sim \sum_{g=1}^G \eta_g f_g(\cdot|\theta_g),
\end{equation}
where different components are allowed different parametric forms; the full generality here is seldom used outside very specific contexts, but cases where the $f_g$ differ only through the values of measured covariates are common; see Section \ref{sec:gen}. Of course, such an assumption modifies the degree of exchangeability in the indexing of components.

Many of the ideas in mixture modelling, including all those raised in this chapter, apply equally to univariate and multivariate data; in the latter case, observed variates $y_i$ are random vectors. Figure \ref{fig:bivnorm} illustrates a bivariate Normal mixture with 2 components. Mixture models are most commonly adopted for continuously distributed variables, but discrete mixtures are also important; an example is illustrated in Figure \ref{fig:poismix}. We continue to use $f$ for what will now be a probability mass function (or density function with respect to counting measure).

Finally, mixture models can equally well be defined as convex combinations of distribution functions or measures; this is little more than a notational variation, and we will use the customary density function representation (\ref{eq:fmm}) throughout this chapter.

\subsection{Likelihood}
Modern inference for mixture models -- whether Bayesian or not -- almost always uses the likelihood function, which in the case of $n$ i.i.d. observations from (\ref{eq:fmm}) has the form
\begin{equation}\label{eq:likd}
p(\ym| \theta,G)= \prod_{i=1}^n \sum_{g=1}^G \eta_g f(y_i|\theta_g).
\end{equation}
Most of the unusual, and sometimes challenging, features of mixture analysis stem from having to handle this product-of-sums form.

\subsection{Latent allocation variables} \label{suballocations}
An alternative starting point for setting up a mixture model is more probabilistic: suppose the population from which we are sampling is heterogeneous: there are multiple groups, indexed by $g=1,2,\ldots,G$, present in the population in proportions $\eta_g,g=1,2,\ldots,G$. When sampling from group $g$, observations are assumed drawn from density $f(\cdot|\theta_g)$. Then we can imagine that an observation $y$ drawn from the population is realised in two steps: first the group $z$ is drawn from the index set $g=1,2,\ldots,G$, with $P(z=g)=\eta_g$, and secondly, given $z$, $y$ is drawn from $f(\cdot|\theta_{z})$.

This two-stage sampling gives exactly the same model (\ref{eq:fmm}) for the distribution of each $y_i$. In a random sample, we have:
\begin{equation}\label{eq:latent}
y_i|z_i \sim f(\cdot|\theta_{z_i}) \quad \text{with} \quad P(z_i=g)=\eta_g,
\end{equation}
independently for $i=1,2,\ldots,n$. (In the case corresponding to (\ref{eq:fmmg}), $f(\cdot|\theta_{z_i})$ would be replaced by $f_{z_i}(\cdot|\theta_{z_i})$.) The $z_i$ are \emph{latent} (unobserved) random variables; they are usually called \emph{allocation variables} in the mixture model context. This construction is illustrated in Figure \ref{fig:latent}, for the same 3-component univariate Normal mixture as in Figure \ref{fig:norm3}.

This alternative perspective on mixture modelling provides a valuable `synthetic' view of the models, complementary to the `analytic' view in (\ref{eq:fmm}).  It is useful to be able to keep both of these perspectives, irrespective of whether the groups being modelled by the different components $g=1,2,\ldots,G$ have any reality (as substantively interpretable subgroups) in the underlying population, or not. In particular, very commonly the latent variable representation (\ref{eq:latent}) is key in computing inference for mixture models, as we will see below, and later in the volume.

This duality of view does bring a peril, that it can be rather tempting to attempt to deliver inference about the allocation variables $z_i$ even in contexts where the mixture model (\ref{eq:fmm}) is adopted purely as a convenient analytic representation.

More abstractly, the synthetic view of mixture modelling provides a prototype example of a much broader class of statistical models featuring latent variables; mixtures provide a useful exemplar for testing out methodological innovation aimed at this broader class.

\cite{titterington:smith:makov:1985} draw a clear distinction between what they call `direct applications' of mixture modelling, where the allocation variables represent real subgroups in the population and (\ref{eq:latent}) is a natural starting point, and `indirect applications' where the components are more like basis functions in a linear representation and we begin with (\ref{eq:fmm}).

A different way to think about the same distinction is more focussed on outcomes. Direct applications are based intrinsically on an assumption of a heterogeneous population, and delivering inference about that heterogeneity is a probable goal, while indirect applications essentially amount to a form of semi-parametric density estimation.

\subsection{A little history}

The creation-myth of mixture modelling involves two of the great intellectuals of the late 19th century, the biometrician, statistician and eugenicist Karl Pearson and the evolutionary biologist Walter Weldon. The latter speculated in 1893 that the asymmetry he observed in a histogram of forehead to body length ratios in female shore crab populations could indicate evolutionary divergence. \cite{pearson:1894} fitted a
univariate mixture of two Normals to Weldon's data by a method of moments: choosing the five parameters of the mixture so that the empirical moments matched that of the model.

However, there are numerous earlier uses of the idea of modelling data using mixtures can be found in the work of \cite{holmes:1892} on measures of wealth disparity, \cite{newcomb:1886} who wanted a model for outliers, and \cite{quetelet:1852}.

\section{Generalisations}\label{sec:gen}

The basic mixture model (\ref{eq:fmm}) sets the scene and provides a template for a rich variety of models, many covered elsewhere in this book. So this section is in part an annotated contents list for the later chapters in the volume.

\subsection{Infinite mixtures}
It is by no means logically necessary or inevitable that the number of components $G$ in (\ref{eq:fmm}) be finite. Countably infinite mixtures
\begin{equation}\label{eq:imm}
y\sim \sum_{g=1}^\infty \eta_g f(\cdot|\theta_g)
\end{equation}
are perfectly well-defined and easily specified, although in practice seldom used. Adopting such a model avoids various difficulties to do with choosing $G$, whether pre-fixed in advance, or inferred from data (see Chapters 4, 6, 7, 8 and 10). And of course, using a \emph{population} model in which the number of components is unlimited in no way implies that  a \emph{sample} uses more than a finite number. Whether $G$ is finite or infinite, there will in general be `empty components', components $g$ for which $z_i\neq g$ for all $i=1,2,\ldots,n$.

The kind of countably infinite mixture that has been most important in applications is the Dirichlet Process mixture \citep{lo:1984}, and its relatives, that form a central methodology in Bayesian nonparametric modelling (see Chapter 6). In Dirichlet process mixtures, the allocation variables $z_i$ are no longer drawn (conditionally) independently from a distribution $(\eta_g)$, but rather jointly from a P\'olya urn process: ties among the $z_i, i=1,2,\ldots,n$ determine a clustering of the indices $1,2,\ldots,n$, in which the probability of any given
set of clusters is given by
$$
\frac{\alpha^d \prod_j (n_j-1)!}{\alpha(\alpha+1)\cdots(\alpha+n-1)}
$$
where $d$ is the number of non-empty clusters, whose sizes are $n_1,n_2,\ldots,n_d\geq 1$, and $\alpha>0$ is a parameter controlling the degree of clustering. This is not the usual approach to defining the Dirichlet Process mixture model, further exposed in Chapters 6 and 17, but \cite{green:richardson:2001} show that it is equivalent to the standard definitions. They also explain how this model arises as a limit of finite mixture models (\ref{eq:fmm}) with appropriate priors on $\{\eta_g,\theta_g\}$, namely where the $\theta_g$ are a priori i.i.d., and $(\eta_1,\eta_2,\ldots,\eta_G)\sim \mathcal{D}(\alpha/G,\alpha/G,\ldots,\alpha/G)$, considered in the limit as $G\to \infty$.

One of the reasons that this is sometimes called a `nonparametric' mixture model is that the complexity of the model automatically adapts to the sample size. In a Dirichlet process mixture model applied to $n$ i.i.d. data, the prior mean number of components is $\sim \alpha \log(n/\alpha)$ for large $n$. However, this cannot be directly compared to a finite mixture model where $G$ is fixed, since that model allows `empty components', that is components in the model to which in a particular sample, no observation is allocated; in the Dirichlet process mixture, only non-empty components are counted.

\subsection{Continuous mixtures}

Another approach to infinity in the number of components starts from the integral form (\ref{eq:cmm}). Whereas the finite mixture model corresponds to the case where $H$ is discrete, in general, $H(\cdot)$ can be \emph{any} probability distribution on the parameter space $\Theta$, and the resulting model remains well-defined.

Continuous mixtures are sometimes called \emph{compound probability distributions}, they are, if you like, distributions with random parameters. One context where these models arise, perhaps disguised, is in simple Bayesian analysis, where the mixing distribution corresponds to the prior, and the base/component distribution to the likelihood.

Several standard models fall into this class; these include exponential family models mixed by their conjugate `priors', in Bayesian language, so giving an explicit unconditional density.

\begin{figure}[h!]
\begin{center}
\resizebox{0.9\textwidth}{!}{\includegraphics{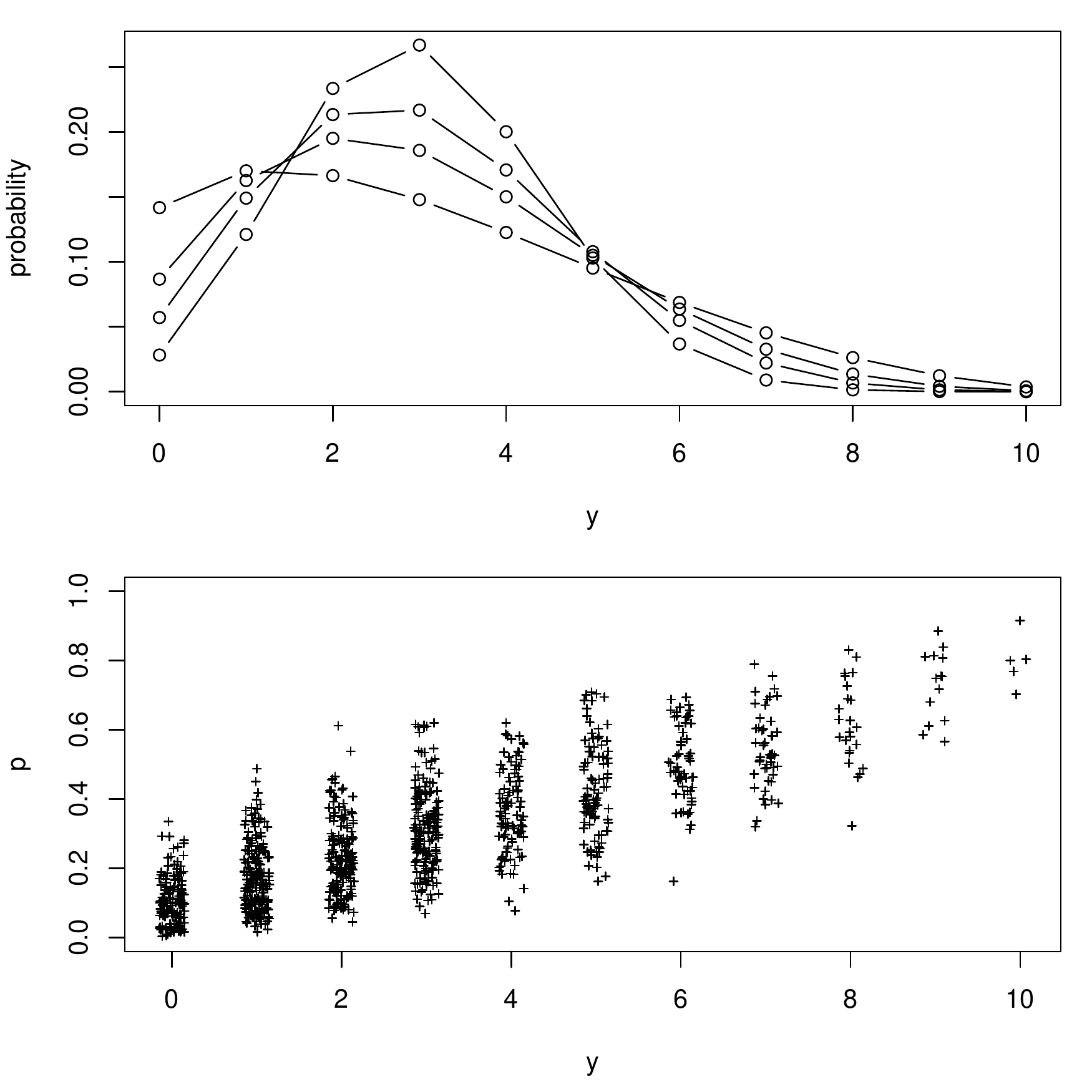}}
\end{center}
\caption{Upper panel: A $\Bino(10,0.3)$ distribution, and beta-binomial distributions with $(a,b)=(6,14)$, $(3,7)$ and $(1.5,3.5)$ (in decreasing order of the height of the mode).
Lower panel: 1000 draws from the distribution of $(y,p)$, where $p\sim \Betadis(1.5,3.5)$ and $y|p\sim \Bino(10,p)$; the points are jittered randomly in the horizontal direction to avoid overplotting.}\label{fig:betabinom}
\end{figure}
A Beta mixture of binomial distributions (each with the same index parameter) is a beta--binomial distribution, often used for data over-dispersed relative to the binomial.
$$
f(y) = \int_0^1 \binom{n}{y} \theta^y (1-\theta)^{n-y} \times
\frac{\theta^{\alpha-1} (1-\theta)^{\beta-1}}{B(\alpha,\beta)} d\theta
= \binom{n}{y} \frac{B(\alpha+y,\beta+n-y)}{B(\alpha,\beta)}
$$
for $y=0,1,2,\ldots,n$. See Figure \ref{fig:betabinom}. When there are more than two categories of response, this becomes a Dirichlet--multinomial distribution.

A Gamma mixture of Poisson distributions is a negative-binomial distribution
$$
f(y) = \int e^{-\theta} \frac{\theta^y}{y!} \times \frac{\theta^{\alpha-1}\beta^\alpha e^{-\beta\theta}}{\Gamma(\alpha)} d\theta = \frac{\Gamma(\alpha+y)}{y!\Gamma(\alpha)} p^y (1-p)^\alpha
$$
for $y=0,1,2,\ldots$, where $p=(1+\beta)^{-1}$, and again a common use is for over-dispersion.

For continuous responses, a commonly-used continuous mixture is the scale mixture of Normals model,
$$
f(y) = \int \phi(y|\mu,\sigma) H(d\sigma)
$$
which preserves the symmetry about $\mu$ of the density, but varies the shape and tail behaviour according to $H$. Familiar examples include the $t$ and double-exponential (Laplace) distributions, which arise when the mixing distribution $H$ is inverse Gamma or exponential respectively.

Another use of continuous mixtures for continuous distributions is provided by the nice observation that any distribution on $[0,\infty)$ with non-increasing density can be represented as a mixture over $\theta$ of uniform distributions on $[0,\theta]$; see for example, \citet[p. 158]{Feller:1970}.

Aspects of continuous mixtures are covered in detail in Chapter~10.

\subsection{Finite mixtures with nonparametric components}

A different kind of mixture modelling with nonparametric aspects other than the Bayesian nonparametric modelling mentioned above takes the mixture model in the form of (\ref{eq:fmmg}), or its alternative representation in terms of distribution functions, and replaces one or more of the components by a nonparametric density, perhaps subject to qualitative constraints. Some examples include the mixture of a known density with an unknown density assumed only to be symmetric, or non-decreasing. Another would be an arbitrary finite mixture of copies of the same unknown parametric density with different unknown location shifts.

Most of the research work on such models has focussed on theoretical questions such as identifiability and rates of convergence in estimation, rather than methodology or application. These models are considered in detail in Chapter~12.

\subsection{Covariates and mixtures of experts}

In regression modelling with mixtures, we want to use analogues of (\ref{eq:fmm}) to model the conditional distribution of a response $y$ given covariates $x$ (scalar or vector, discrete or continuous, in any combination). There are many ways to do this: the weights $\eta_g$ could become $\eta_g(x)$, and/or the component densities could become $f(\cdot|\theta_g,x)$ or $f(\cdot|\theta_g(x))$, and these $\eta_g(x)$ and $\theta_g(x)$ could be parametric or nonparametric regression functions, as the modeller requires.

The case where both $\eta_g(x)$ and $\theta_g(x)$ appear as functions of the covariate,
$$
y|x \sim \sum_{g=1}^G \eta_g(x) f(\cdot|\theta_g(x)),
$$
is known in the machine learning community as a \emph{mixture of experts} model \citep{jacobs:jordan:1991}, the component densities $f(\cdot|\theta_g(x))$ being the `experts', and the weights $\eta_g(x)$ the `gating networks'. The special cases where only $\eta_g$ or only $\theta_g$ depend on $x$ have been called the `gating-network' and the `expert-network' mixture of experts models.

Mixtures of experts models are fully discussed in Chapter 13.

\subsection{Hidden Markov models}
\begin{figure}[ht]
\begin{center}
\resizebox{0.9\textwidth}{!}{\includegraphics{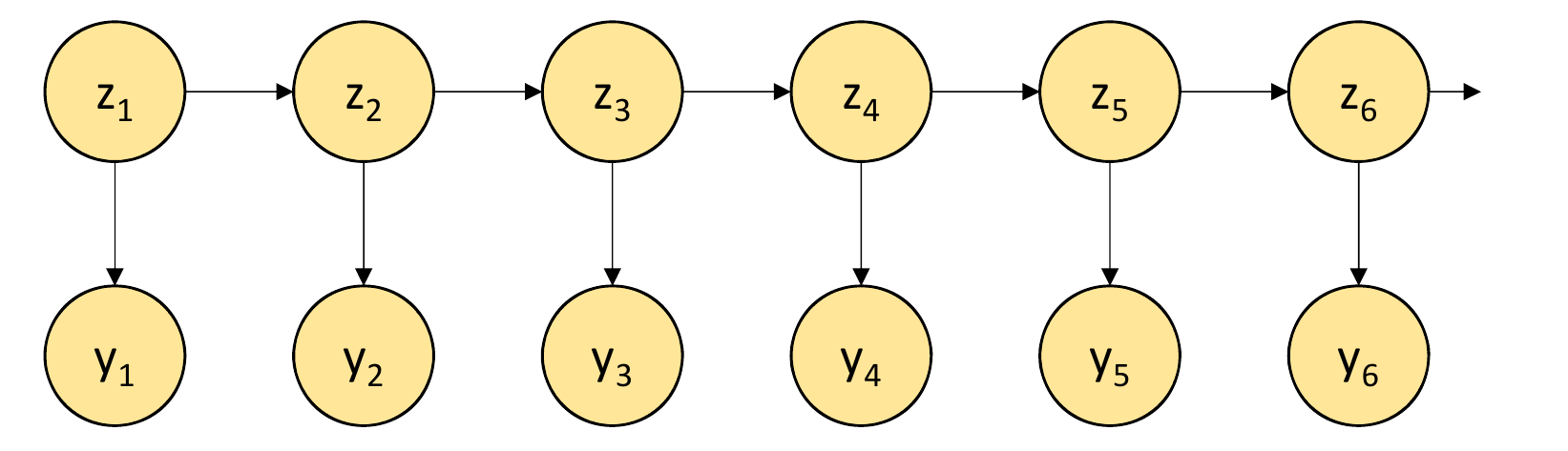}}
\end{center}
\caption{Graphical representation of a hidden Markov model (as a directed acyclic graph).}\label{fig:hmmdag}
\end{figure}
In contexts where responses $y_t,t=1,2,\ldots,T$ are delivered sequentially in time, it is natural to consider mixture models analogous to (\ref{eq:fmm}) but capturing sequential dependence. This is most conveniently modelled by supposing that the allocation variables, now denoted $z_t$, form a (finite) Markov chain, say on the states $\{1,2,\ldots,G\}$ and with transition matrix $\xi$. This defines a \emph{Hidden Markov model}
\begin{gather}\label{eq:hmm}
y_t|z_1,z_2,\ldots,z_t \sim f(\cdot|\theta_{z_t}) \\
P(z_t=h|z_1,z_2,\ldots, z_{t-1}=g) = \xi_{gh} \nonumber
\end{gather}
Figure \ref{fig:hmmdag} displays the structure of this model as a directed acyclic graph. This set-up can naturally and dually be interpreted either as a finite mixture model with serial dependence, or as a Markov chain observed only indirectly (as the word `hidden' suggests) through the `emission distributions' $f(\cdot|\theta_{z_t})$. It is also a template for a rich variety of models extending the idea to (countably) infinite state spaces, and to continuous state spaces, where these are usually called \emph{state-space} models. Also very important in many applications are hidden Markov models where the order of dependence in the state sequence $z_1,z_2,\ldots, z_t,\ldots$ is greater than 1 (or of variable order).

Hidden Markov models, and other time-series models involving mixtures, are fully discussed in Chapter 11 and Chapter 17.

\subsection{Spatial mixtures}

Where data are indexed spatially rather than temporally, mixture models can play an analogous role, and as usual the models proposed can often be viewed as generalisations or translations from time to space. For problems where the observational units are discrete locations or non-overlapping regions of space, the Hidden Markov random field model of \cite{ge02} was introduced as an analogue of (\ref{eq:hmm}) where the serial Markov dependence among $z_t$ is replaced by a Potts model, a discrete Markov random field, with $t$ now indexing locations or regions in space. Letting $\zm=\{z_t\}$ be the collection of all unknown allocations, then we take
$$
p(\zm) \propto \exp(\psi\sum_{t'\sim t} \mathbb{I}(z_t=z_{t'})),
$$
where $t'\sim t$ denotes that spatial locations $t$ and $t'$ are contiguous in some sense, and
$\psi$ is an interaction parameter, positive for positive association. In the context of \cite{ge02}, the observations $y_t$ were counts of cases of a rare disease in regions indexed by $t$, modelled as conditionally independent given $\zm$, with
$$
y_t \sim \mathcal{P}(\theta_{z_t}E_t),
$$
where $(\theta_g)_{g=1}^G$ are parameters, a priori i.i.d., and $(E_t)$ are expected counts based on population size, adjusted for, say, age and sex.
Again, there are dual interpretations, as a mixture model with hidden spatial dependence, or as a spatial process observed with noise. In the latter view, this model can then be associated with the rich variety of image models built on hidden spatial Markov processes for the `true scene' \citep{geman:1984,besag:1986}.

An alternative to modelling dependence among the allocation variables directly through a discrete stochastic process is to introduce spatial dependence between the \emph{distributions} of the allocation variables, that is between the weights at different spatial locations. Models of this kind were introduced by \cite{fernandez:green:2002dplctd}, which assumes conditionally-independent spatially-indexed observations $y_t$ are modelled as
$$
y_t \sim \sum_{g=1}^G \eta_{tg} f_t(\cdot|\theta_g)
$$
where the (location-specific) weights $\eta_{tg}$ are modelled as a spatial stochastic process;
they have something of the flavour of (gating-network) mixtures of expert models where the 'covariate' is spatial location. \citet{fernandez:green:2002dplctd} consider two specific forms for  the spatial weights $\eta_{tg}$: one a multinomial logit transformation of (latent) independent Gaussian processes, the other using a grouped-continuous model, also based on a Gaussian process.
For the same rare-disease mapping context, the component densities were given by
$$
f_t(y|\theta_g)= e^{-\theta_gE_t} \frac{(\theta_gE_t)^y}{y!}
$$
Spatial mixtures, and in particular their application to image analysis, are discussed in Chapter 16.

\section{Some Technical Concerns} \label{sec:tech}

\subsection{Identifiability}

\cite{fruhwirth:2006} observes that the density (\ref{eq:fmm}) remains invariant on either (a) including an additional component with arbitrary parameter $\theta$ but weight $\eta=0$, or (b) replicating any of the components, giving the new one(s) the same parameter $\theta$ and sharing with it the existing weight $\eta$, and that this violates identifiability. In commenting on this observation, \citet{mengersen:rousseau:2011} write
``This non-identifiability is much stronger than the non-identifiability corresponding to permutations of the labels in the mixture representation''. I would go further and argue that the label-permutation invariance is best not thought of as non-identifiability at all.

Leaving aside special situations where the mixture representation (\ref{eq:fmm}) is adopted in a situation where the component labels $g$ represent real pre-identified groups in a population -- and where, for example a Bayesian analysis would use a prior specification in which different priors could be used for different groups -- the model is indeed invariant to permutation of the labels $g=1,2,\ldots,G$. But it seems to me to be wrong to refer to this as lack of identifiability, it is not a property of the mixture model (\ref{eq:fmm}), but a side-effect of the representation used therein.

When the mixing distribution $H$ has finite support, the mixture model (\ref{eq:fmm}) is equivalent to the integral representation (\ref{eq:cmm}). The only difference is that in (\ref{eq:fmm}) the atoms of the distribution $H$ have been numbered, $g=1,2,\ldots$. But there is no natural or canonical way to do this numbering: the labels $g$ are artificial creatures of the choice of representation, and serve only to pair the parameter values $\theta_g$ with their corresponding weights $\eta_g$.

If we focus, more properly, on inference for the mixing distribution $H$, then identifiability is possible: for example, \citet{teicher:1963} showed that finite mixtures of Normal or Gamma distributions are identifiable (but interestingly this is not true for binomial distributions). Focussing on $H$ does more than solve the problem of invariance to label-permutation, it even fixes the `stronger' non-identifiability that is obtained by allowing zero-weight or coincident components, since such variants do not affect the mixing distribution $H$ either.

The objective of \citet{mengersen:rousseau:2011} is to study the impact of this `non-identifiability' on asymptotic inference; it would be interesting to see whether the difficulties they uncover also apply in the case of asymptotic inference about $H$.

\subsection{Label switching}

A related but different problem, but one with the same root cause, is the phenomenon of `label-switching'. This has attracted a lot of attention from researchers and generated many papers. It shows up particularly in iterative methods of inference for a finite mixture model, especially Markov chain Monte Carlo methods for Bayesian mixture analysis. However the problem can be stated, and solved, without even mentioning iteration.

Consider the model (\ref{eq:fmm}), with estimates $\widehat{\theta}_g$ and $\widehat{\eta}_g$ of the unknown quantities. Then not only is the model (\ref{eq:fmm}) invariant to permutation of the labels of components, so is the \emph{estimated model}
\begin{equation}\label{eq:fmmest}
y\sim \sum_{g=1}^G \widehat{\eta_g} f(\cdot|\widehat{\theta}_g).
\end{equation}
Given that both numberings of components are arbitrary, there is no reason whatsoever for the numbering of the components in (\ref{eq:fmm}) and (\ref{eq:fmmest}) to be coherent. For example, we cannot say that $\widehat{\theta}_1$ estimates $\theta_1$!

In a MCMC run, the lack of coherence between the numberings can apply separately at each iteration, so commonly `switchings' are observed.

Numerous solutions to this perceived problem have been proposed, but the only universal valid and appropriate approach to both apparent non-identifiability and label-switching is to recognise that you can only make inference about the mixing distribution $H$. You can estimate (or find the posterior distributions of) the number of (distinct) atoms of $H$ that lie in a given set, or the total weight of these atoms, or the weight of the component with the largest variance, or indeed any other property of the mixture distribution $\sum_{g=1}^G \eta_g f(\cdot|\theta_g)$ but you cannot unambiguously estimate, say, $\theta_1$ or $\eta_3$.

If the group labels do have a substantive meaning, for example, group $g=1$ is `the group with the largest variance', then $\theta_1$ and $\eta_1$ do become individually meaningful. But you still cannot expect this substantive meaning to automatically attach to the labelling of the groups in the \emph{estimated} mixture $\sum_{g=1}^G \widehat{\eta_g} f(\cdot|\widehat{\theta_g})$.

\section{Inference}

This opening chapter is not the place to go into inference for mixture models in any detail, but this section attempts to give the flavour of some of the algorithmic ideas for basic inference in finite mixture models.

\subsection{Frequentist inference, and the role of EM}

An enormous variety of methods has been considered over the years for frequentist inference about unknowns in a mixture model. For the standard problem of estimation based on a simple random sample from (\ref{eq:fmm}), Pearson's oft-cited work \citep{pearson:1894} used the method of moments to estimate the 5 free parameters in a two-component univariate Normal mixture, and for at least a century thereafter novel contributions to such methods for mixtures continued to appear. A second big group of methods is based on a suitable distance measure between the empirical distribution of the data sample and the mixture model, to be minimised over choice of the free parameters.
Except for some very particular special cases, both the method of moments and minimum-distance methods require numerical optimisation, there being no explicit formulae available.

In current practice the dominant approach to inference is by maximum likelihood, based on maximisation of (\ref{eq:likd}), again using numerical methods. Subject to solving the numerical issues, such an approach is immediately appealing on many grounds, including its continuing applicability when the basic model is elaborated, through the introduction of covariates for example.

The usual numerical approach to maximum likelihood estimation of mixture models, advocated for example throughout the book by \citet{maclachlan01}, uses the EM algorithm \citep{demp:lair:rubi:1977}. Mixture modelling and the EM algorithm are identified together so strongly that mixtures are often used as illustrative examples in tutorial accounts of EM. This is highly natural, not only because of statisticians' typical conservatism in their toolkits of numerical methods, but because the natural area of applicability of EM, to `hidden data' problems, maps so cleanly onto the mixture model set-up.

Referring back to the latent allocation variable formulation (\ref{eq:latent}), it is clear that if the $z_i$ were known, we would have separate independent samples for each component $g$, so that estimation of $\theta_g$ would be a standard task. Conversely, if the parameters $\theta_g$ were known, allocating observations $y_i$ to the different components could be done naturally by choosing $z_i=g$ to maximise $f(y_i|\theta_g)$. One could imagine a crude algorithm that alternates between these two situations; this could be seen as a `decision-directed' or `hard classification' version of the EM algorithm, as applied to mixtures.

The actual EM algorithm is the `soft-classification' analogue of this. The $z_i$ are treated as the `missing data', and represented by the indicator vectors with components $z_{ig}$, where $(z_{ig}=1) \equiv (z_i=g)$. In the E-step of the algorithm $z_{ig}$ is replaced by its conditional expectation given the current values of the parameters and weights, that is
\begin{equation}\label{eq:estep}
z_{ig} \leftarrow \frac{\eta_g f(y_i|\theta_g)}{\sum_{g'} \eta_{g'}f(y_i|\theta_{g'})},
\end{equation}
and on the M-step, $\eta_g$ and $\theta_g$ are updated to maximise the corresponding expected complete-data log-likelihood
$$
\sum_i \sum_g z_{ig} \log(\eta_g f(y_i|\theta_g)),
$$
which can often be done in closed form. When the $\eta_g$ and $\theta_g$ vary independently for each component $g$, this maximisation may be done separately for each. The EM algorithm, though sometimes slow, enjoys the usual advantages that each iteration increases, and under certain conditions maximises, the likelihood.

We see that in this algorithm, the latent allocation variables $z_i$ play a crucial role, irrespective of whether the groups have a substantive interpretation.

Maximum likelihood inference for mixture models is covered in full detail in Chapters 2 and 3.

\subsection{Bayesian inference, and the role of MCMC}\label{sec:bayesinf}

This elevation of the allocation variables to essential ingredients of a numerical method is also seen in the most natural MCMC method for Bayesian inference of the parameters and weights. \citet{die} and \citet{dieb:robe:1994} proposed data augmentation and Gibbs sampling approaches to posterior sampling for parameters and weights of finite mixture distributions. Given the current values of $\{\eta_g\}$ and $\{\theta_g\}$, the Gibbs sampler update for the allocation variables $z_i$ is to draw them independently from $P(z_i=g)=z_{ig}$, where $z_{ig}$ is as in (\ref{eq:estep}); this is simply Bayes' rule. Note the close parallel with EM here. 

MCMC updating of the $\eta_g$ and $\theta_g$ requires Gibbs sampling, or more generally Metropolis-Hastings updates, using the full conditionals of these variables given the data $\ym$ and the current values of the allocation variables $\zm$. The details here depend on the form of the base density $f$ and the priors on the parameters; for Normal mixtures under Normal--inverse Gamma priors as in \citet{dieb:robe:1994}, the Gibbs sampling update is straightforward and involves only sampling from Dirichlet, inverse Gamma and Normal densities.

There is an obvious parallel between the EM algorithm for the maximum likelihood solution and the MCMC algorithm for posterior sampling in the conjugate-prior Bayesian model, with the same kind of central role for the latent allocation variables.

Bayesian inference for mixture models is discussed in detail in Chapter 4, while Chapter 5 is devoted to  posterior sampling.

\subsection{Variable number of components}
One of the many appeals of a Bayesian approach to mixture analysis is that it supports a particularly clean way of dealing with uncertainty in the number $G$ of components. In a Bayesian set-up, inference about $G$ can be performed simultaneously with that about the weights $\eta_g$ and component parameters $\theta_g$, whereas in frequentist inference, $G$ has to be treated as a model indicator, an unknown of a different character than the other unknowns, the traditional parameters. For the rest of this subsection, we consider only the Bayesian approach.

When $G$ is not treated as known we are in one of those situations where `the number of unknowns is one of the things you don't know', a type of problem that \citet{roeder-wasserman-1997} termed `transdimensional'. The objective of the inference will be to deliver the joint posterior distribution of $\{G,(\eta_g,\theta_g)_{g=1}^G\}$. There are numerous possible approaches to the computational task of computing this posterior, of which sampling-based approaches are much the most popular. As often happens with MCMC methods for transdimensional problems, those for mixture analysis fall into two categories: within-model and across-model samplers.

Within-model sampling involves running separate MCMC posterior simulations for each of a range of values of $G$ from 1 up to some pre-assigned maximum of interest. Each individual simulation is a standard fixed-$G$ analysis, along the lines of Section \ref{sec:bayesinf}. In addition to delivering a fixed-$G$ posterior sample, the simulation must calculate, or rather estimate, the marginal likelihood $p(\ym|G)$ of the data, given $G$. After all runs are complete, the separate inferences can be combined across $G$ using the fact that
$$
p(G|\ym) = \frac{p(G) p(\ym|G)}{\sum_G p(G) p(\ym|G)}.
$$

In across-model sampling, a single MCMC sampler is constructed, targeting the joint posterior of $\{G,(\eta_g,\theta_g)_{g=1}^G\}$, using for example the method of \citet{carlin:chib:1995}, reversible jump MCMC \citep{greenetrichardson97} or a birth-and-death process as in \citet{stephens:2000}, see Chapter~7.

Both within- and across-model simulations can be used to deliver exactly the same conclusions, since both ultimately estimate the same joint posterior in a simulation-consistent way. One or another approach may be more computationally efficient for the same precision of a particular quantity of interest, depending on circumstances.

\begin{figure}[ht]
\begin{center}
\resizebox{0.9\textwidth}{!}{\includegraphics{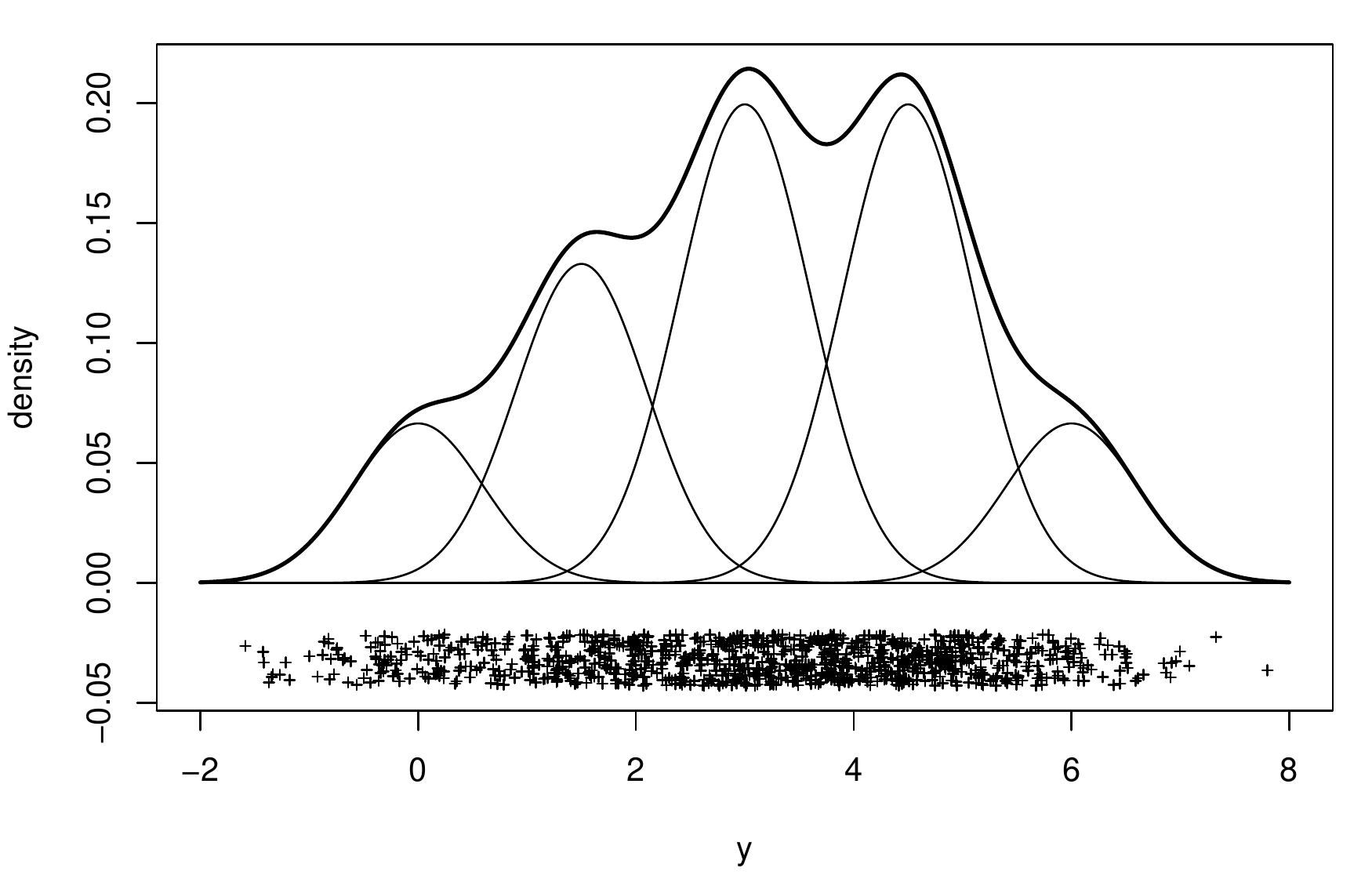}}
\end{center}
\caption{A Normal mixture model with 5 components, 3 modes, and for which a sample of 1000 draws seem to form 1 cluster.
The model is $0.1\times \Normal(0,0.6^2)+0.2\times \Normal(1.5,0.6^2)+0.3\times \Normal(3, 0.6^2)+0.3\times \Normal(4.5,0.6^2)+0.1\times \Normal(6,0.6^2)$.
}\label{fig:modeclus}
\end{figure}

\subsection{Modes versus components}
Motivation for mixture modelling often stems from visual exploratory data analysis, where a histogram of data suggests multimodality. But of course there need be no simple relationship between the number of modes in a mixture such as (\ref{eq:fmm}) and the number $G$ of components. Indeed, Fig. \ref{fig:norm3} displays a three-component mixture with only one mode, while there is nothing in (\ref{eq:fmm}) to prevent even a single component having any number of modes. If component densities in (\ref{eq:fmm}) are unimodal, then the mixture will have $G$ modes provided the $\theta_g$ are sufficiently different.

Rigorous statistical theory and methodology taking explicit note of modality is rare, though recent work by \citet{chacon:2015} provides a population background for this.
\citet{amendola:2017} discuss the history of the problem of enumerating the possible numbers of modes for a mixture of $G$ Gaussian distributions in $d$ dimensions, and gives improved lower
bounds, and the first upper bound, on the maximum number of non-degenerate modes.

\subsection{Clustering and classification}
In a rather similar way, mixture models may be adopted under an informal belief that the data are clustered, with observations within each cluster drawn from some simple parametric distribution. It is a small step apparently to regard `sub-populations' as `clusters', but again there may be no simple relationship between inferred components in a fitted mixture model and clusters in the original data. The term `cluster' commonly suggests both a degree of \emph{homogeneity} within a cluster and a degree of \emph{separation} between clusters, while typically \emph{not} implying any specific within-cluster distribution. A fitted mixture model may need more components that there are apparent clusters in the data simply because the assumed parametric form of the component densities is incorrect, so that each homogeneous cluster requires multiple components to fit it well. See the artificial example in Figure \ref{fig:modeclus}.

Nevertheless, finite mixture modelling does provide one of the few simple and rigorous model-based approaches to clustering, a theme that is explored in detail in Chapter 8.

For convincing model-based inference about clusters, using a within-cluster data distribution that is itself likely to be a mixture, it seems necessary to adopt a second level of indexing of weights and parameters, as in
$$
y\sim \sum_{g=1}^G \sum_{h=1}^{H_g} \eta_{gh} f(\cdot|\theta_{gh}),
$$
where $g$ indexes clusters and $h$ components within clusters. 
The demands of homogeneity within clusters and separation between clusters can be met by appropriately modeling the $\theta_{gh}$; for example, in a Bayesian formulation, we might require $\theta_{gh}-\theta_{g'h'}$ to have small variance if $g=g'$ and large variance otherwise, perhaps most conveniently achieved with a hierarchical formulation such as $\theta_{gh}=\alpha_g+\beta_{gh}$, with a suitably `repelling' prior on the $\alpha_g$. A more intricate version of this basic idea has been thoroughly investigated recently in \citet{mal-etal:ide}.

\section{Conclusion}

In this opening chapter, I hope to have set the scene on some key ideas of mixture modelling, stressing the underlying unity of the main concepts, as they apply to different data structures and in different contexts. The full versatility of mixtures as a modelling tool, the wide range of specific methodologies, the rich and subtle theoretical underpinnings of inference in the area, and ideas for further research will only become fully clear as the rest of the book unfolds.


\begin{thebibliography}{}

\bibitem[\protect\citename{Am{\'{e}}ndola {\em et~al.}, }2017]{amendola:2017}
Am{\'{e}}ndola, Carlos, Engstr{\"{o}}m, Alexander, \& Haase, Christian. 2017.
\newblock {\em Maximum Number of Modes of {G}aussian Mixtures}.
\newblock arXiv preprint arXiv:1702.05066.

\bibitem[\protect\citename{Besag, }1986]{besag:1986}
Besag, J. 1986.
\newblock On the statistical analysis of dirty pictures.
\newblock {\em J. Royal Statist. Society Series B}, {\bf 48}, 259--279.

\bibitem[\protect\citename{Carlin \& Chib, }1995]{carlin:chib:1995}
Carlin, B.P., \& Chib, S. 1995.
\newblock {B}ayesian model choice through {M}arkov chain {M}onte {C}arlo.
\newblock {\em J. Royal Statist. Society Series B}, {\bf 57}(3), 473--484.

\bibitem[\protect\citename{Chac\'on, }2015]{chacon:2015}
Chac\'on, Jos\'e~E. 2015.
\newblock A Population Background for Nonparametric Density-Based Clustering.
\newblock {\em Statistical Science}, {\bf 30}(4), 518--532.

\bibitem[\protect\citename{Dempster {\em et~al.}, }1977]{demp:lair:rubi:1977}
Dempster, Arthur~P., Laird, Nan~M., \& Rubin, Donald~B. 1977.
\newblock Maximum Likelihood from Incomplete Data via the {EM} Algorithm (with
  discussion).
\newblock {\em J. Royal Statist. Society Series B}, {\bf 39}, 1--38.

\bibitem[\protect\citename{Diebolt \& Robert, }1994]{dieb:robe:1994}
Diebolt, J., \& Robert, Christian~P. 1994.
\newblock Estimation of Finite Mixture Distributions by {B}ayesian Sampling.
\newblock {\em J. Royal Statist. Society Series B}, {\bf 56}, 363--375.

\bibitem[\protect\citename{Diebolt \& Robert, }1990]{die}
Diebolt, J., \& Robert, {C.P.} 1990.
\newblock Estimation des param{\`e}tres d'un m{\'e}lange par
  {\'e}chantillonnage bay{\'e}sien.
\newblock {\em Notes aux Comptes--Rendus de l'Acad{\'e}mie des Sciences I},
  {\bf 311}, 653--658.

\bibitem[\protect\citename{Feller, }1970]{Feller:1970}
Feller, W. 1970.
\newblock {\em An Introduction to Probability Theory and its Applications}.
\newblock  Vol. 1.
\newblock New York: John Wiley.

\bibitem[\protect\citename{Fern\'andez \& Green,
  }2002]{fernandez:green:2002dplctd}
Fern\'andez, C., \& Green, P.~J. 2002.
\newblock Modelling spatially correlated data via mixtures: a {B}ayesian
  approach.
\newblock {\em Journal of the Royal Statistical Society Series B}, {\bf 64},
  805--826.

\bibitem[\protect\citename{Fr{\"u}hwirth-Schnatter, }2006]{fruhwirth:2006}
Fr{\"u}hwirth-Schnatter, S. 2006.
\newblock {\em Finite Mixture and Markov Switching Models}.
\newblock New York: Springer-Verlag, New York.

\bibitem[\protect\citename{Geman \& Geman, }1984]{geman:1984}
Geman, S., \& Geman, D. 1984.
\newblock Stochastic relaxation, {G}ibbs distributions and the {B}ayesian
  restoration of images.
\newblock {\em IEEE Trans. Pattern Anal. Mach. Intell.}, {\bf 6}, 721--741.

\bibitem[\protect\citename{Green \& Richardson, }2001]{green:richardson:2001}
Green, P.J., \& Richardson, S. 2001.
\newblock Modelling heterogeneity with and without the {D}irichlet process.
\newblock {\em Scandinavian J.~Statistics}, {\bf 28}(2), 355--375.

\bibitem[\protect\citename{Green \& Richardson, }2002]{ge02}
Green, P.J., \& Richardson, S. 2002.
\newblock Hidden {M}arkov models and disease mapping.
\newblock {\em J. American Statist. Assoc.}, {\bf 92}, 1055--1070.

\bibitem[\protect\citename{Holmes, }1892]{holmes:1892}
Holmes, G.~K. 1892.
\newblock Measures of distribution.
\newblock {\em Journal of the American Statistical Association}, {\bf 3},
  141--157.

\bibitem[\protect\citename{Jacobs {\em et~al.}, }1991]{jacobs:jordan:1991}
Jacobs, R.A., Jordan, M.I., Nowlan, S.~J., \& Hinton, G.~E. 1991.
\newblock Adaptive mixture of local experts.
\newblock {\em Neural Computation}, {\bf 3}, 19--87.

\bibitem[\protect\citename{Lo, }1984]{lo:1984}
Lo, Albert~Y. 1984.
\newblock On a class of {B}ayesian nonparametric estimates: I. {D}ensity
  estimates.
\newblock {\em The Annals of Statistics}, {\bf 12}(1), 351--357.

\bibitem[\protect\citename{Malsiner-Walli {\em et~al.}, }2017]{mal-etal:ide}
Malsiner-Walli, Gertraud, Fr\"uhwirth-Schnatter, Sylvia, \& Gr\"un, Bettina.
  2017.
\newblock Identifying Mixtures of Mixtures Using {B}ayesian Estimation.
\newblock {\em Journal of Computational and Graphical Statistics}, {\bf 26}(2),
  285--295.

\bibitem[\protect\citename{McLachlan \& Peel, }2000]{maclachlan01}
McLachlan, G.~J., \& Peel, D. 2000.
\newblock {\em Finite {M}ixture {M}odels}.
\newblock New York: J. Wiley.

\bibitem[\protect\citename{Newcomb, }1886]{newcomb:1886}
Newcomb, Simon. 1886.
\newblock A generalized theory of the combination of observations so as to
  obtain the best result.
\newblock {\em American Journal of Mathematics}, {\bf 8}(4), 343--366.

\bibitem[\protect\citename{Pearson, }1894]{pearson:1894}
Pearson, K. 1894.
\newblock Contribution to the mathematical theory of evolution.
\newblock {\em Proc. Trans. Royal Society A}, {\bf 185}, 71--110.

\bibitem[\protect\citename{Quetelet, }1852]{quetelet:1852}
Quetelet, A. 1852.
\newblock Sur quelques propriet\'es curieuses que pr\'esentent les re\'sultats
  d'une s\'erie d'observations, faites dans la vue de d\'eterminer une
  constante, lorsque les chances de rencontrer des \'ecarts en plus et en moins
  sont \'egales et ind\'ependant\'es les unes des autres.
\newblock {\em Bulletin de l'Acad\'emie Royale des Sciences, des Lettres et des
  Beaux-Arts de Belgique}, {\bf 19}, 303--317.

\bibitem[\protect\citename{Richardson \& Green, }1997]{greenetrichardson97}
Richardson, Sylvia, \& Green, Peter~J. 1997.
\newblock On {B}ayesian analysis of mixtures with an unknown number of
  components.
\newblock {\em J. Roy. Statist. Soc. Ser. B}, {\bf 59}(4), 731--792.

\bibitem[\protect\citename{Roeder \& Wasserman, }1997]{roeder-wasserman-1997}
Roeder, Kathryn, \& Wasserman, Larry. 1997.
\newblock Contribution to the discussion of the paper by {R}ichardson and
  {G}reen.
\newblock {\em Journal of the Royal Statistical Society Series B}, {\bf 59}(4),
  782.

\bibitem[\protect\citename{Rousseau \& Mengersen,
  }2011]{mengersen:rousseau:2011}
Rousseau, Judith, \& Mengersen, Kerrie. 2011.
\newblock Asymptotic behaviour of the posterior distribution in overfitted
  mixture models.
\newblock {\em J. Royal Statist. Society Series B}, {\bf 73}(5), 689--710.

\bibitem[\protect\citename{Stephens, }2000]{stephens:2000}
Stephens, M. 2000.
\newblock {B}ayesian analysis of mixture models with an unknown number of
  components---an alternative to reversible jump methods.
\newblock {\em Ann. Statist.}, {\bf 28}, 40--74.

\bibitem[\protect\citename{Teicher, }1963]{teicher:1963}
Teicher, H. 1963.
\newblock Identifiability of finite mixtures.
\newblock {\em Ann. Math. Statist.}, {\bf 34}, 1265--1269.

\bibitem[\protect\citename{Titterington {\em et~al.},
  }1985]{titterington:smith:makov:1985}
Titterington, D.M., Smith, A.F.M., \& Makov, U.E. 1985.
\newblock {\em {S}tatistical Analysis of Finite Mixture Distributions}.
\newblock New York: John Wiley.

\end{thebibliography}
\hyphenation{Post-Script Sprin-ger}

\end{document}